\documentclass[twocolumn,english,aps,pra,showpacs,superscriptaddress]{revtex4-2}
\usepackage[latin9]{inputenc}
\usepackage{verbatim}
\usepackage{amsmath}
\usepackage{amssymb}
\usepackage{graphicx}
\usepackage{esint}

\makeatletter

\providecommand{\tabularnewline}{\\}

\makeatother

\usepackage{babel}
\begin{document}
\title{Photon pair antibunching and second-order correlations between pair
events}
\author{Chien-Chang Chen}
\affiliation{Department of Physics and Center for Theoretical Physics, National
Taiwan University, Taipei 106319, Taiwan}
\affiliation{Center for Quantum Science and Engineering, National Taiwan University,
Taipei 106319, Taiwan}
\author{Ite A. Yu}
\affiliation{Department of Physics and Center for Quantum Science and Technology,
National Tsing Hua University, Hsinchu 30013, Taiwan}
\affiliation{National Center for Excellence in Quantum Information Science and
Engineering, National Tsing Hua University, Hsinchu 30013, Taiwan}
\author{Hsi-Sheng Goan}
\email{goan@phys.ntu.edu.tw}

\affiliation{Department of Physics and Center for Theoretical Physics, National
Taiwan University, Taipei 106319, Taiwan}
\affiliation{Center for Quantum Science and Engineering, National Taiwan University,
Taipei 106319, Taiwan}
\affiliation{Physics Division, National Center for Theoretical Sciences, Taipei
106319, Taiwan}
\date{\today}
\begin{abstract}
We introduce the pair second-order correlation function $g_{\textrm{pairs}}^{\left(2\right)}=\left\langle \left(P^{\dagger}\right)^{2}P^{2}\right\rangle /\left\langle P^{\dagger}P\right\rangle ^{2}$,
defined through the pair operator $P^{\dagger}=a^{\dagger}b^{\dagger}$,
to characterize second-order correlations and pair bunching and antibunching
in photon-pair creation processes. This quantity directly probes correlations
between pair-generation events within a single two-mode quantum state,
providing access to the intrinsic pair-generation process beyond conventional
single-mode or heralded second-order correlations, which do not directly
capture correlations between pair events. Values of $g_{\textrm{pairs}}^{\left(2\right)}$
greater than, equal to, or less than unity correspond respectively
to pair bunching, Poissonian pair statistics, and pair antibunching.
Using the Cauchy--Schwarz inequality, we further show that all classical
two-mode fields described by a positive Glauber--Sudarshan \ensuremath{P}-function
satisfy $g_{\textrm{pairs}}^{\left(2\right)}\geq1$, so that pair
antibunching is classically forbidden and constitutes an unambiguous
signature of nonclassicality. We evaluate $g_{\textrm{pairs}}^{\left(2\right)}$
for several representative quantum states and show, in particular,
that even arbitrarily weak two-mode squeezed vacuum states exhibit
pair bunching. Comparison with heralded second-order correlations
highlights the complementary information provided by these observables.
The proposed correlation function is experimentally accessible via
standard coincidence measurements, requires no phase reference or
state reconstruction, and remains invariant under uniform loss.
\end{abstract}
\maketitle

\section*{INTRODUCTION}

Nonclassical correlations in biphoton sources underpin a wide range
of quantum technologies, including heralded single-photon generation,
quantum communication, and entanglement distribution. Traditional
nonclassicality criteria such as Wigner-function \citep{Milburn08}.
negativity or quantum Fisher information (QFI) associated with a chosen
observable provide powerful diagnostics, but their evaluation typically
requires state reconstruction or phase-sensitive measurements. Heralded
single-mode intensity correlations $g_{s}^{\left(2\right)}$ probe
conditional single-photon statistics and provide a widely used signature
of antibunching \citep{Antibunching_Kimble1977,antibunching_quantum_Mandel1979,Click_g2_eta_i_Horoshko2019}.
However, they characterize only marginal single-mode properties and
do not directly access the statistics of the underlying pair-generation
process. The unconditional signal--idler cross-correlation $g_{si}^{\left(2\right)}$,
extensively employed to quantify intermode intensity correlations
in biphoton sources \citep{cross_g2_Bocquillon2009,Ite_experiment_g2_2024},
characterizes correlations between photon numbers in the two modes
but does not directly access the statistics of the pair-generation
operator itself.

While single-photon antibunching has long served as a hallmark of
nonclassical light, an analogous antibunching criterion for genuine
biphoton emission has not been established. In this paper, we introduce
such a criterion through pair-level correlation measurements. A more
faithful indicator of pairwise quantum behavior should directly probe
the statistics of the pair (biphoton) creation operator $P^{\dagger}=a^{\dagger}b^{\dagger}$
\footnote{The corresponding annihilation operator $P=ab$ does not satisfy canonical
bosonic commutation relations, i.e., $\left[P,P^{\dagger}\right]\neq1$,
and therefore does not admit a conventional bosonic number-operator
interpretation.}. Such an operator-level perspective captures the statistics of the
pair-generation process itself, rather than the marginal statistics
of either optical mode. Despite their formal simplicity, pair-resolved
correlation functions built from $P^{\dagger}P$ have received relatively
little attention as practical nonclassicality witnesses, in part because
most existing criteria are formulated in terms of single-mode photon-number
moments or quadrature distributions \citep{Normal_ordering_Glauber1963,quadrature_Vogel2000}.
Establishing a rigorous, experimentally accessible, and loss-tolerant
benchmark for genuinely quantum pair emission therefore is a natural
and important goal.

\section*{RESULTS}

Here we show that the pair second-order correlation function

\begin{eqnarray}
g_{\textrm{pairs}}^{\left(2\right)} & = & \frac{\left\langle P^{\dagger}P^{\dagger}PP\right\rangle }{\left\langle P^{\dagger}P\right\rangle ^{2}},\label{eq:g2pairs}
\end{eqnarray}
provides exactly such a benchmark and reveals a distinct regime of
pair correlations inaccessible to classical fields. Using the Cauchy--Schwarz
inequality, we prove that all classical two-mode fields with a positive
Glauber--Sudarshan \ensuremath{P}-function \citep{P_function_Glauber1963}
satisfy the universal bound $g_{\textrm{pairs}}^{\left(2\right)}\geq1$.
Observation of antibunching at the level of pair correlations, $g_{\textrm{pairs}}^{\left(2\right)}<1$,
is therefore classically forbidden and defines a genuinely nonclassical
regime of photon-pair emission, directly tied to statistics of the
biphoton operator. As with other normally ordered intensity correlations,
$g_{\textrm{pairs}}^{\left(2\right)}$ is invariant under standard
linear (beam-splitter) loss \citep{Milburn08}, since both numerator
and denominator scale proportionally with detection efficiency and
thus cancel. In contrast, heralded correlations depend explicitly
on the conditioning efficiency. These properties make pair-level antibunching
an experimentally accessible and robust signature of nonclassical
biphoton emission that can be measured through standard coincidence-counting
techniques, without phase stabilization or full state reconstruction.

\subsection*{Theorem Proof}

We now establish the classical bound for the pair correlation function
defined in Eq. (\ref{eq:g2pairs}). For a classical two-mode field
with a positive Glauber--Sudarshan \ensuremath{P}-function $P\left(\alpha,\beta\right)\geq0$,
all normally ordered moments can be expressed as averages over complex
field amplitudes $\alpha$ and $\beta$,
\begin{equation}
\left\langle :f\left(a^{\dagger},a,b^{\dagger},b\right):\right\rangle =\int P\left(\alpha,\beta\right)f\left(\alpha^{\dagger},\alpha,\beta^{\dagger},\beta\right)d^{2}\alpha d^{2}\beta,\label{eq:Prep}
\end{equation}
where $f\left(a^{\dagger},a,b^{\dagger},b\right)$ is any normally
ordered function of the operators.

Under this representation, the expectation values appearing in Eq.
(\ref{eq:g2pairs}) reduce to moments of the classical pair amplitude
$z=\alpha\beta$. Specifically,
\begin{equation}
\langle P^{\dagger}P\rangle=\langle|z|^{2}\rangle,\qquad\langle P^{\dagger}P^{\dagger}PP\rangle=\langle|z|^{4}\rangle,\label{eq:zmap}
\end{equation}
where the averages are taken with respect to the positive probability
distribution $P\left(\alpha,\beta\right)$. The pair correlation function
can therefore be written as
\begin{eqnarray}
g_{\textrm{pairs}}^{\left(2\right)} & = & \frac{\left\langle \left|z\right|^{4}\right\rangle }{\left\langle \left|z\right|^{2}\right\rangle ^{2}}.\label{eq:g2classical}
\end{eqnarray}
Applying the Cauchy--Schwarz inequality to the classical random variable
$\left|z\right|^{2}$,
\begin{eqnarray}
\left\langle \left|z\right|^{4}\right\rangle =\left\langle \left|z\right|^{2}\cdot\left|z\right|^{2}\right\rangle  & \geq & \left\langle \left|z\right|^{2}\right\rangle ^{2},\label{eq:CS}
\end{eqnarray}
we immediately obtain the universal bound
\begin{equation}
g_{\textrm{pairs}}^{\left(2\right)}\geq1.\label{eq:classicalbound}
\end{equation}
Equation (\ref{eq:classicalbound}) holds for all classical two-mode
fields with a positive Glauber--Sudarshan \ensuremath{P}-function.
Because linear loss preserves classicality, the bound remains valid
under arbitrary linear attenuation (e.g., transmission loss or detection
inefficiency). Any observation of pair antibunching, $g_{\textrm{pairs}}^{\left(2\right)}<1$
, therefore implies the absence of a positive Glauber--Sudarshan
\ensuremath{P}-function and constitutes a genuine signature of nonclassical
pair generation.

\subsection*{Examples of pair source}

We now illustrate the behavior of the pair correlation function $g_{\textrm{pairs}}^{\left(2\right)}$
for representative classical and quantum two-mode states. 

(i) Two-mode coherent state. 

For the product coherent state $\left|\alpha\right\rangle \otimes\left|\beta\right\rangle $,
all normally ordered moments factorize. The biphoton operator therefore
obeys Poissonian statistics, yielding $g_{\textrm{pairs}}^{\left(2\right)}=1$,
which saturates the classical bound.

(ii) Two-mode squeezed vacuum (TMSV). 

The TMSV state is given by \citep{TMSVGuha2009,TMSV_Valivarthi2020,squeezing_amplitude2022,TMSVBrougham2023}
\begin{equation}
|\psi_{\mathrm{TMSV}}\rangle=\sqrt{1-\lambda^{2}}\sum_{n=0}^{\infty}\lambda^{n}|n,n\rangle,\label{eq:TMSV}
\end{equation}
where $\lambda=\tanh r$ and $r$ denotes the two-mode squeezing amplitude
\citep{Click_g2_eta_i_Horoshko2019,squeezing_amplitude2022}. This
state exhibits perfect photon-number correlations between the two
modes and serve as a paradigmatic entangled biphoton source. Direct
evaluation shows $g_{\textrm{pairs}}^{\left(2\right)}>1$ for all
$\lambda\neq0$, demonstrating pair bunching despite perfect photon-number
correlations between the two modes. The explicit derivation is provided
in Appendix \ref{sec:appendix_pair_g2_TMSV}.

(iii) Pair-antibunched and truncated TMSV sources. 

As the opposite extreme, consider the truncated state
\begin{equation}
|\psi_{\mathrm{AB}}\rangle=\sqrt{1-p}|0,0\rangle+\sqrt{p}|1,1\rangle,\label{eq:Pair-antibunched}
\end{equation}
where $p$ is the probability of generating a single photon pair,
which may be regarded as the lowest truncation order ($N=1$) of the
TMSV expansion. Because the two-pair probability vanishes identically,
one obtains $g_{\textrm{pairs}}^{\left(2\right)}=0$, representing
maximal pair antibunching and the strongest possible violation of
the classical bound. Although this state coincides algebraically with
the first-order weak-squeezing expansion of the TMSV, it is treated
here as a physically distinct source in which the multipair sector
is explicitly removed.

\begin{figure}
\includegraphics[width=1\columnwidth]{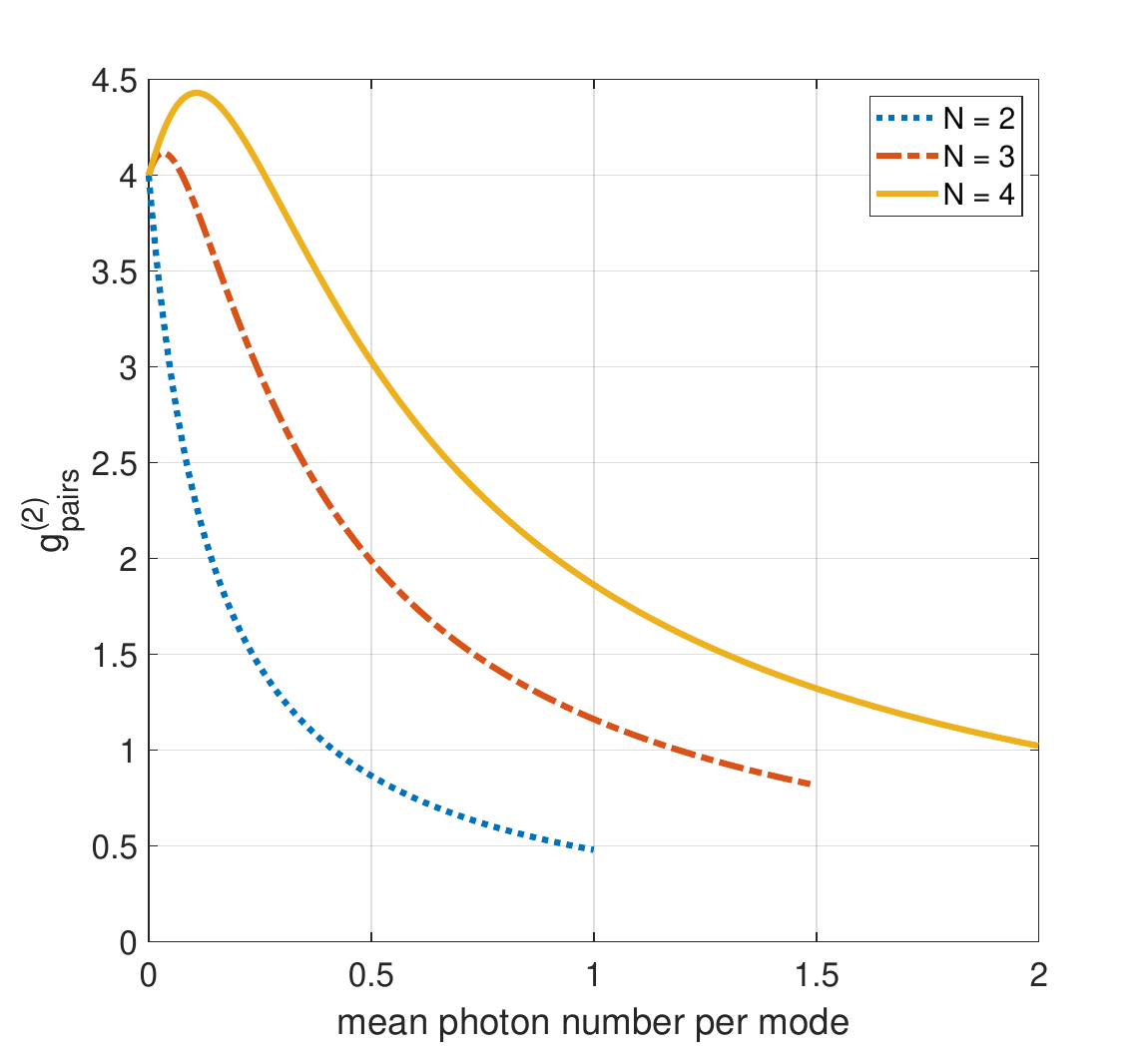}\caption{Pair second-order correlation $g_{\textrm{pairs}}^{\left(2\right)}$
as a function of the mean photon number per mode $\bar{n}$ for truncated
TMSV states with truncation orders $N=2$,3, and 4. All cases approach
the universal weak-squeezing limit $g_{\textrm{pairs}}^{\left(2\right)}\rightarrow4$
at low $\bar{n}$. For $N=2$ and $N=3$, the pair correlation exhibits
a crossover from bunching to antibunching as $\bar{n}$ increases,
while for $N=4$ the statistics remain bunched over the entire parameter
range. The curves are obtained from Eq. (\ref{eq:pair_g2_N_truncated_TMSV}).}
\label{fig:pair_N2_truncated_TMSV}
\end{figure}

More generally, finite-order truncations of Eq. (\ref{eq:TMSV}) with
$N\geq2$ share the same weak-squeezing limiting value $g_{\textrm{pairs}}^{\left(2\right)}\rightarrow4$,
since, for any $N\geq2$, the one- and two-pair sectors universally
dominate as $\lambda\rightarrow0$. However, their large-squeezing
behavior depends strongly on the truncation order. As shown in Fig.
\ref{fig:pair_N2_truncated_TMSV}, for the lowest truncation orders
$N=2$ and $N=3$, the pair correlation exhibits a crossover from
bunching to antibunching as $\lambda\rightarrow1$, whereas for higher
truncation orders (e.g., $N=4$) the statistics remain bunched over
the entire parameter range. Each truncation order corresponds to a
bounded range of accessible mean photon number $\bar{n}$, reflecting
the finite maximum pair number in the truncated state. In particular,
the minimal nontrivial $N=2$ case approaches the antibunched limiting
value $g_{\textrm{pairs}}^{\left(2\right)}=12/25<1$, as derived in
Appendix \ref{sec:appendix_N2_truncated_TMSV}.

This behavior contrasts with the more familiar heralded $g_{s}^{\left(2\right)}$,
where increasing brightness typically enhances multi-pair contributions
and leads to stronger bunching. In the present case, however, the
pair correlation is defined in terms of the operator $P=ab$ and depends
on the relative scaling of different pair-number contributions. For
truncated states, increasing $\lambda$ does not open higher-pair
channels but instead redistributes population toward the highest allowed
pair number, reducing the relative weight of pair coincidences and
leading to the observed antibunching.

\section*{DISCUSSION}

It is important to clarify the role of the product coherent state
as the reference for uncorrelated (Poissonian) pair-event statistics
in defining the pair second-order correlation function. In the two-mode
setting, a \textquotedblleft pair event\textquotedblright{} is defined
operationally as a joint detection event in the two modes, independent
of the underlying generation mechanism or the presence of signal--idler
correlations. One might expect that such a reference should itself
be a correlated two-mode (\textquotedblleft pair\textquotedblright )
source, since the quantity of interest concerns pair correlations.
However, the defining property of Poissonian pair statistics is not
the presence of correlations in the source, but rather the statistical
independence of joint detection events. The natural baseline for such
statistics is therefore a state in which detection events in the two
modes are statistically independent.

Within this framework, the condition, $\left\langle P^{\dagger}P^{\dagger}PP\right\rangle =\langle P^{\dagger}P\rangle^{2}$,
i.e., $g_{\textrm{pairs}}^{\left(2\right)}=1$, signifies Poissonian
pair-event statistics corresponding to independent joint detection
events. This condition defines the boundary between pair bunching
and pair antibunching.

The product coherent state satisfies this criterion exactly: although
it does not constitute a correlated \textquotedblleft pair source\textquotedblright ,
it exhibits statistically independent detection events in the two
modes and therefore realizes $g_{\textrm{pairs}}^{\left(2\right)}=1$.
More generally, the classical bound on $g_{\textrm{pairs}}^{\left(2\right)}$
follows from assuming a classical complex pair amplitude $z=\alpha\beta$
with a positive probability distribution and applying the Cauchy--Schwarz
inequality, without requiring any specific physical source or inter-mode
correlation.

Consequently, the pair second-order correlation function $g_{\textrm{pairs}}^{\left(2\right)}$
provides a direct criterion for characterizing pair statistics: values
$g_{\textrm{pairs}}^{\left(2\right)}>1$ correspond to pair bunching,
while $g_{\textrm{pairs}}^{\left(2\right)}<1$ signify pair antibunching
and indicate suppressed multi-pair emission. The Poissonian value
$g_{\textrm{pairs}}^{\left(2\right)}=1$ marks the boundary between
these regimes and the classical threshold; values below unity are
forbidden for classical fields and therefore constitute an unambiguous
signature of nonclassical pair statistics. When observed in a verified
pair source, pair antibunching further indicates strongly suppressed
multi-pair emission, analogous to photon antibunching in single-photon
sources. These distinctions are summarized schematically in Table
\ref{table_classicality_pair_bunching}.

\begin{table}
\begin{tabular}{|c|c|c|}
\hline 
 & Pair bunching  & Pair antibunching\tabularnewline
\hline 
Classical & $g_{\textrm{pairs}}^{\left(2\right)}>1$ & forbidden\tabularnewline
\hline 
Nonclassical & $g_{\textrm{pairs}}^{\left(2\right)}>1$ & $g_{\textrm{pairs}}^{\left(2\right)}<1$\tabularnewline
\hline 
\end{tabular}

\caption{Classification of pair-emission statistics according to classicality
and pair bunching behavior. The pair second-order correlation function
$g_{\textrm{pairs}}^{\left(2\right)}$ distinguishes pair bunching
($g_{\textrm{pairs}}^{\left(2\right)}>1$) from pair antibunching
($g_{\textrm{pairs}}^{\left(2\right)}<1$). Classical fields are restricted
to the bunched or Poissonian regime, while pair antibunching is forbidden
classically and constitutes an unambiguous signature of nonclassical
pair statistics.}

\label{table_classicality_pair_bunching}
\end{table}

\begin{figure}
\includegraphics[width=1\columnwidth]{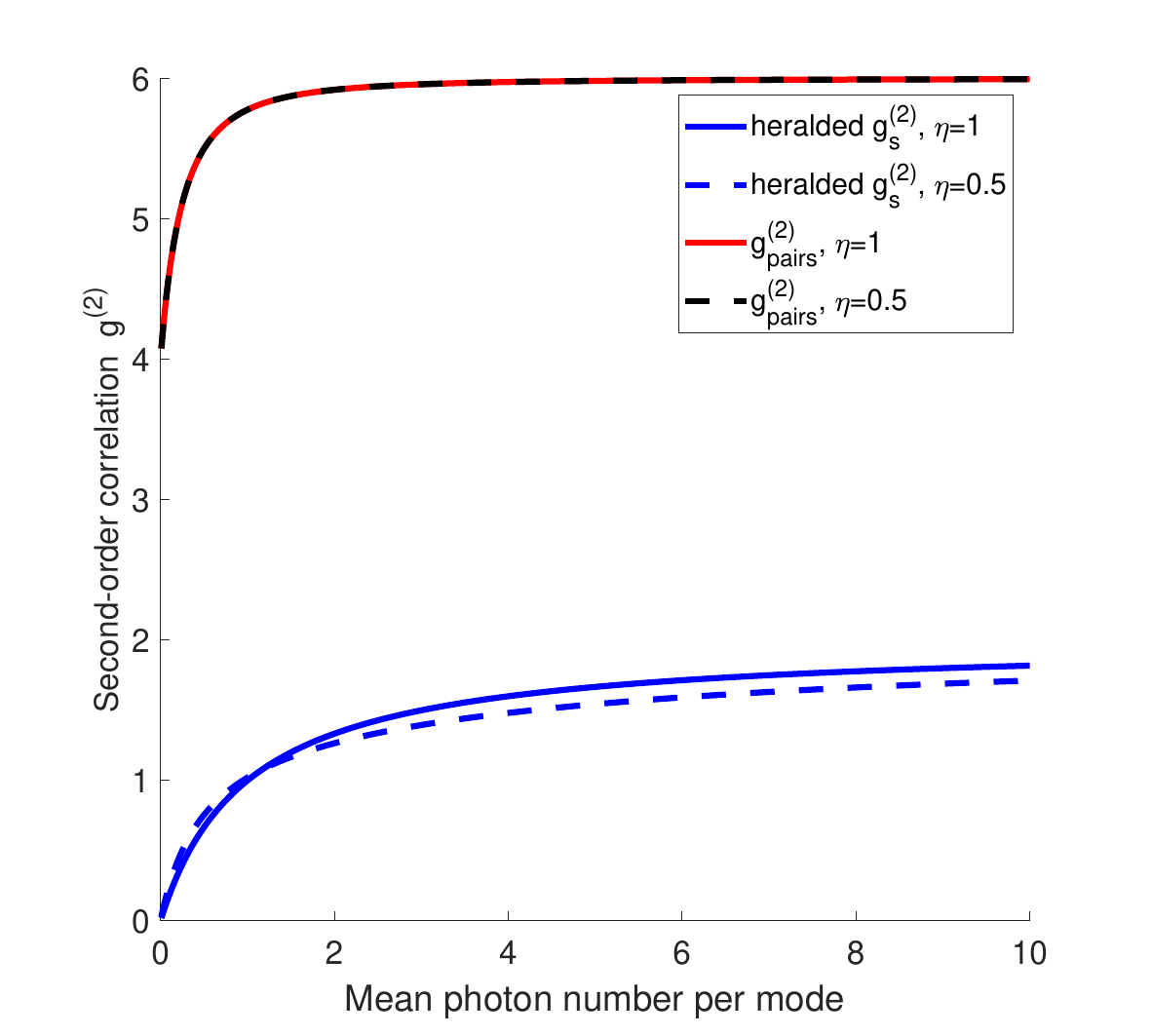} \caption{Comparison of pair and heralded second-order correlations for a two-mode
squeezed vacuum state as a function of the mean photon number per
mode, $\bar{n}=\sinh^{2}r$. Shown are the pair correlation $g_{\textrm{pairs}}^{\left(2\right)}$
and the heralded signal correlation $g_{s}^{\left(2\right)}$ conditioned
on an idler click, plotted for detection efficiencies $\eta=1$ and
$\eta=0.5$. While the heralded correlation can exhibit antibunching
due to conditional post-selection, $g_{\textrm{pairs}}^{\left(2\right)}$
probes intrinsic pair statistics and does not necessarily show antibunching
in the same regime.}
\label{fig:pair_vs_cond_g2}
\end{figure}

Together, these examples demonstrate that the inequality $g_{\textrm{pairs}}^{\left(2\right)}\geq1$
sharply separates classical and quantum regimes, with its violation
identifying nonclassical sources characterized by suppressed second-order
pair statistics. Unlike single-mode or heralded intensity correlations,
the pair correlation directly probes statistics of the total number
of photon pairs generated per realization and therefore accesses a
distinct aspect of nonclassical light. 

This distinction has an important physical consequence: $g_{\textrm{pairs}}^{\left(2\right)}$
need not exhibit antibunching even for genuinely quantum sources.
Heralded $g_{s}^{\left(2\right)}$ characterizes conditional single-photon
statistics and is sensitive to how closely the post-selected state
approximates an ideal single-photon state. By contrast, even when
photons are created strictly in signal--idler pairs, the number of
such pairs can fluctuate significantly from shot to shot, leading
to pair bunching rather than antibunching. Consequently, antibunching
in conditional single-photon statistics does not imply suppressed
second-order pair statistics, and the two witnesses diagnose complementary
properties of the same quantum state. 

This behavior is summarized in Fig. \ref{fig:pair_vs_cond_g2}, which
contrasts intrinsic pair correlations with the heralded auto-correlation
measures for the TMSV state. The figure shows both ideal ($\eta=1$)
and reduced ($\eta=0.5$) idler detection efficiencies. While linear
loss associated with idler detection efficiency quantitatively modifies
the heralded $g_{s}^{\left(2\right)}$, it does not affect the qualitative
presence of antibunching in the weak-pumping regime ($\bar{n}\ll1$),
where heralded experiments typically operate. In contrast, the pair
correlation $g_{\textrm{pairs}}^{\left(2\right)}$ remains invariant
under uniform loss, as indicated by the overlapping curves for $\eta=1$
and $\eta=0.5$. The explicit analytical expressions for the heralded
$g_{s}^{\left(2\right)}$, including linear loss, have been presented
in Ref. \citep{Click_g2_eta_i_Horoshko2019}, and those for the pair
correlation function $g_{\textrm{pairs}}^{\left(2\right)}$ are provided
in the Appendix \ref{sec:appendix_pair_g2_TMSV}.

At a more fundamental level, it is instructive to contrast the single-pair
truncation ($N=1$) with the full TMSV state. Truncating the TMSV
to the \{$n=0,1$\} subspace yields pair antibunching by construction,
as shown in Eq. (\ref{eq:Pair-antibunched}), and is often invoked
under weak-excitation approximations \citep{weak_excitation_TMSV2022}.
However, the exact TMSV state exhibits $g_{\textrm{pairs}}^{\left(2\right)}>4$
for any nonzero squeezing parameter, irrespective of how weak the
source is. This demonstrates that even an arbitrarily small multipair
component---negligible for first-order observables---qualitatively
alters second-order pair correlations.

A related distinction arises when comparing with the more familiar
heralded $g_{s}^{\left(2\right)}$. While reducing the pump strength
suppresses multipair contributions through conditioning and drives
the signal toward a single-photon state, the unconditional TMSV retains
its full multipair structure, and its normalized pair statistics remain
unchanged as the excitation is reduced. Consequently, pair bunching
persists arbitrarily close to the weak-field limit. This highlights
that antibunching cannot be inferred from weak excitation alone, but
requires explicit removal of multipair contributions, as realized
by state truncation.

\subsection*{Experimental Accessibility}

The pair correlation function $g_{\textrm{pairs}}^{\left(2\right)}$
can be accessed experimentally using standard photon-correlation techniques
without requiring phase-sensitive detection or state tomography. Since
$P^{\dagger}P=a^{\dagger}b^{\dagger}ab$ corresponds to joint photon-number
coincidences between signal and idler modes, the numerator $\left\langle P^{\dagger}P^{\dagger}PP\right\rangle $
can be obtained from fourfold coincidence measurements, while the
denominator involves squared twofold coincidences. Such measurements
can be implemented using photon-number-resolving detectors or multiplexed
threshold detectors combined with time-tagged coincidence counting,
as routinely employed in biphoton experiments based on spontaneous
parametric down-conversion or four-wave mixing.

\section*{Conclusion}

In summary, we have identified the pair correlation function $g_{\textrm{pairs}}^{\left(2\right)}$
as a simple and rigorous witness of nonclassical biphoton emission.
Expressed in terms of the biphoton operator $P^{\dagger}=a^{\dagger}b^{\dagger}$,
this quantity directly probes statistics of the pair-generation process.
Using the Cauchy--Schwarz inequality, we establish the universal
classical bound $g_{\textrm{pairs}}^{\left(2\right)}\geq1$ for all
two-mode fields with a positive Glauber--Sudarshan \ensuremath{P}-function,
so that observation of pair antibunching $g_{\textrm{pairs}}^{\left(2\right)}<1$
certifies nonclassicality. We further show that even weakly excited
two-mode squeezed vacuum states exhibit pair bunching, in contrast
to the antibunching behavior commonly observed in heralded single-mode
measurements. This distinction reveals that pair-level and conditional
single-photon statistics probe complementary aspects of quantum light,
with $g_{\textrm{pairs}}^{\left(2\right)}$ providing a direct pair-level
witness of nonclassicality. We note, however, that pair antibunching
certifies nonclassicality of the joint two-mode field rather than
correlated biphoton generation itself, since product states containing
a nonclassical single-mode component may also violate the classical
bound. Moreover, the witness is experimentally accessible using standard
coincidence measurements and is robust against uniform loss, requiring
neither phase reference nor state reconstruction. These features make
pair-level antibunching a practical and transparent diagnostic for
nonclassical light in contemporary biphoton platforms.
\begin{acknowledgments}
H.-S. Goan acknowledges support from the National Science and Technology
Council (NSTC), Taiwan, under Grants No. NSTC 113-2112-M-002-022-MY3,
No. NSTC 113-2119-M-002-021, No. NSTC 114-2119-M-002-018, No. NSTC
114-2119-M-002-017-MY3 and from the National Taiwan University under
Grants No. NTU-CC-115L8937, No. NTU-CC-115L893704 and No. NTU-CC-115L8512.
H.-S. Goan. is also grateful for the support of the \textquotedblleft Center
for Advanced Computing and Imaging in Biomedicine (NTU-115L900702)\textquotedblright{}
through the Featured Areas Research Center Program within the framework
of the Higher Education Sprout Project by the Ministry of Education
(MOE), Taiwan, the support of Taiwan Semiconductor Research Institute
(TSRI) through the Joint Developed Project (JDP) and the support from
the Physics Division, National Center for Theoretical Sciences, Taiwan.

\begin{comment}

\section*{Data Availability}

The numerical data underlying the figures were generated from analytical
expressions presented in this article and from formulas cited from
the literature. No external datasets were used.

\section*{Code Availability}

No custom code was used in this study.

\section*{Competing Interests}

Author Hsi-Sheng Goan serves on the Editorial Boards of EPJ: Quantum
Technology, the International Journal of Quantum Information, and
the Chinese Journal of Physics. Hsi-Sheng Goan was not involved in
the journal\textquoteright s review of, or decisions related to, this
manuscript. The other authors declare no competing financial or non-financial
interests.

\section*{Author contributions}

C.-C.C. conceived and developed the theoretical framework, performed
the analytical calculations, and drafted the manuscript. H.-S.G. supervised
the research, contributed to the interpretation of the results, and
revised the manuscript. I.A.Y. contributed to the original motivation
of the work, its physical interpretation, and its experimental relevance,
and revised the manuscript. All authors discussed the results and
approved the final manuscript.
\end{comment}
\end{acknowledgments}

\appendix

\section{$g_{\textrm{pairs}}^{\left(2\right)}$ of a Two-Mode Squeezed Vacuum}

\label{sec:appendix_pair_g2_TMSV}

We omit operator hats for notational simplicity.

For the two-mode squeezed vacuum (TMSV) defined in the main text with
$\lambda=\tanh r$, we express all results in terms of the mean photon
number per mode, $\bar{n}=\sinh^{2}r.$

\subsection*{Mean pair (number) intensity}

We first evaluate the expectation value
\begin{equation}
\left\langle P^{\dagger}P\right\rangle =\left\langle a^{\dagger}b^{\dagger}ab\right\rangle .
\end{equation}
For the TMSV, which is a zero-mean Gaussian state, Gaussian moment
factoring (Wick\textquoteright s theorem) yields
\begin{equation}
\left\langle a^{\dagger}b^{\dagger}ab\right\rangle =\left\langle a^{\dagger}a\right\rangle \left\langle b^{\dagger}b\right\rangle +\left\langle ab\right\rangle \left\langle a^{\dagger}b^{\dagger}\right\rangle .
\end{equation}
Using $\left\langle a^{\dagger}a\right\rangle =\left\langle b^{\dagger}b\right\rangle =\bar{n}$
and
\begin{eqnarray}
\left\langle ab\right\rangle  & = & e^{i\varphi}\cosh r\sinh r\equiv m,\\
\left\langle a^{\dagger}b^{\dagger}\right\rangle  & = & m^{*},
\end{eqnarray}
with $\left|m\right|^{2}=\bar{n}\left(\bar{n}+1\right)$, we obtain
\begin{equation}
\left\langle P^{\dagger}P\right\rangle =\bar{n}\left(2\bar{n}+1\right).
\end{equation}

\subsection*{Second-order (eight-operator) expectation value}

We now evaluate

\begin{equation}
\left\langle P^{\dagger}P^{\dagger}PP\right\rangle =\left\langle a^{\dagger}b^{\dagger}a^{\dagger}b^{\dagger}abab\right\rangle ,
\end{equation}
which involves eight field operators.

Similarly, using Gaussian moment factoring, this expectation value
can be reduced to combinations of second-order correlations of the
TMSV. Evaluating all contractions yields
\begin{equation}
\left\langle P^{\dagger}P^{\dagger}PP\right\rangle =4\bar{n}^{2}\left(6\bar{n}^{2}+6\bar{n}+1\right).
\end{equation}

\subsection*{Pair correlation function}

Combining the above results, we obtain
\begin{eqnarray}
g_{\textrm{pairs}}^{\left(2\right)} & = & \frac{4\left(6\bar{n}^{2}+6\bar{n}+1\right)}{\left(2\bar{n}+1\right)^{2}}.
\end{eqnarray}
In the limits of small and large mean photon number, $g_{\textrm{pairs}}^{\left(2\right)}\rightarrow4$
for $\bar{n}\rightarrow0$ and $g_{\textrm{pairs}}^{\left(2\right)}\rightarrow6$
for $\bar{n}\rightarrow\infty$, as plotted in Fig \ref{fig:pair_vs_cond_g2}.

\section{General finite-order truncated TMSV expression and the $N=2$ example}

\label{sec:appendix_N2_truncated_TMSV}

As an explicit example of the finite-order truncation discussed in
the main text, we consider the TMSV truncated at order $N$,
\begin{equation}
\left|\psi_{\leq N}\right\rangle =\frac{\sum_{n=0}^{N}\lambda^{n}\left|n,n\right\rangle }{\sqrt{\sum_{n=0}^{N}\lambda^{2n}}}.\label{eq:N_truncated_TMSV}
\end{equation}
Here $0\leq\lambda<1$ follows the parameter range convention of the
physical TMSV. For the full TMSV, the parameter is related to the
two-mode squeezing amplitude $r$ through $\lambda=\tanh r$. For
the truncated TMSV in Eq. (\ref{eq:N_truncated_TMSV}), however, $\lambda$
should be regarded simply as a state parameter characterizing the
relative weights of different pair-number components. For compactness,
we define $x=\lambda^{2}$, thus $0\leq x<1$. The limits $x\ll1$
and $x\rightarrow1$ will be referred to as the weak- and large-squeezing
limits, respectively.

For the pair annihilation operator $P=ab$, the pair second-order
correlation function takes the general form
\begin{equation}
g_{\textrm{pairs}}^{\left(2\right)}=\frac{\left\langle \left(P^{\dagger}\right)^{2}P^{2}\right\rangle }{\left\langle P^{\dagger}P\right\rangle ^{2}}=\frac{\left[\sum_{n=0}^{N}n^{2}\left(n-1\right)^{2}x^{n}\right]\left(\sum_{n=0}^{N}x^{n}\right)}{\left(\sum_{n=0}^{N}n^{2}x^{n}\right)^{2}}.\label{eq:pair_g2_N_truncated_TMSV}
\end{equation}
In the weak-squeezing limit $x\ll1$, all finite-order truncations
with $N\geq2$ are dominated by the vacuum, one-pair, and two-pair
sectors, while higher-pair terms are suppressed by higher powers of
$x$. To leading order, $\sum_{n=0}^{N}x^{n}\approx1+x$, while
\begin{eqnarray}
\sum_{n=0}^{N}n^{2}x^{n} & \approx & x+4x^{2},\\
\sum_{n=0}^{N}n^{2}\left(n-1\right)^{2}x^{n} & \approx & 4x^{2},
\end{eqnarray}
yielding $g_{\textrm{pairs}}^{\left(2\right)}\rightarrow4$. Thus
all finite-order truncations with $N\geq2$ share the same weak-squeezing
behavior.

As the minimal nontrivial explicit example, we now consider the $N=2$
truncation,
\begin{equation}
|\psi_{\leq2}\rangle=\frac{\left|0,0\right\rangle +\lambda\left|1,1\right\rangle +\lambda^{2}\left|2,2\right\rangle }{\sqrt{1+\lambda^{2}+\lambda^{4}}}.
\end{equation}
The relevant pair moments are
\begin{eqnarray}
\langle P^{\dagger}P\rangle & = & \frac{x+4x^{2}}{1+x+x^{2}},\\
\langle\left(P^{\dagger}\right)^{2}P^{2}\rangle & = & \frac{4x^{2}}{1+x+x^{2}},
\end{eqnarray}
which gives
\begin{eqnarray}
g_{\textrm{pairs}}^{\left(2\right)} & = & \frac{4\left(1+x+x^{2}\right)}{\left(1+4x\right)^{2}}.
\end{eqnarray}
In the large-squeezing limit $x\rightarrow1$, the pair correlation
approaches $g_{\textrm{pairs}}^{\left(2\right)}=12/25<1,$ explicitly
demonstrating asymptotic pair antibunching for this minimal nontrivial
truncation. The corresponding mean photon number per mode is
\begin{equation}
\bar{n}=\left\langle \psi_{\leq2}\right|a^{\dagger}a\left|\psi_{\leq2}\right\rangle =\frac{x+2x^{2}}{1+x+x^{2}},
\end{equation}
which in the same limit approaches $\bar{n}\rightarrow1$.

\bibliographystyle{apsrev4-2}
\bibliography{Pair_Antibunching_witness_reference}

\end{document}